\documentclass[prl,twocolumn,showpacs]{revtex4}
\usepackage{graphicx}
\usepackage{amsmath}
\usepackage{amsfonts}
\usepackage{mathrsfs}

\newcommand{\ps}{\ps}

\newcommand{\Up}{{\hat{\mathcal U}}}

\newcommand{\Np}{{\hat{\mathcal N}}}

\newcommand{\Ub}{\Up_{\beta}}

\begin{document}
\title{Ballistic and localized transport for the atom optics kicked rotor
in the limit of vanishing kicking period}
\author{Mark Sadgrove,$^1$ Sandro Wimberger,$^{2}$ Scott Parkins,$^1$ Rainer Leonhardt$^1$}
\affiliation{$^1$Department of Physics, University of Auckland, Private Bag 92019,
Auckland, New Zealand \\ 
$^2$Dipartimento di Fisica E. Fermi, Universit\`{a} di Pisa,
Largo Pontecorvo 3, 56127 Pisa, Italy}
\begin{abstract}
We present mean energy measurements for the atom optics kicked rotor 
as the kicking period tends to zero. A narrow resonance
is observed  marked by quadratic energy growth, in parallel with
a complete freezing of the energy absorption away from the resonance peak. 
Both phenomena are explained by classical means, taking proper account
of the atoms' initial momentum distribution.
\end{abstract}
\maketitle
\date{today}

The atom optics realization of the paradigmatic kicked rotor (KR) \cite{CCFI79}
allows the experimental study of uniquely quantum mechanical aspects 
of a fundamental, classically nonlinear system. Dynamical localization is
perhaps the most celebrated quantum phenomenon observed in the quantum KR \cite{CCFI79,Shepelyansky1987}, 
however the phenomenon of quantum resonance, whereby some atoms experience enhanced energy growth 
(quadratic in kick number) for well-defined kicking periods, has 
received renewed theoretical and experimental attention of late.

The possibility for quadratic energy growth to occur for the KR at quantum resonance has been
known for some time \cite{Izr}. The first experiments to observe phenomena related to
quantum resonance employed the atom optics kicked rotor (AOKR) to study the momentum distributions
of ensembles of kicked rotors \cite{Raizen}, although the broad initial momentum distributions involved
prohibited the observation of quadratic mean energy growth. Recent 
studies have shown
that the quantum resonances, initially thought to be extremely sensitive quantum effects,
can be explained by purely classical means \cite{Wimberger2003}, and that, indeed,
only linear growth in mean energy is expected at 
quantum resonance for experiments starting with
a broad initial momentum distribution \cite{Wimberger2003}. 
The experimental signatures of quantum resonance behaviour
prove to be robust with respect to decoherence by spontaneous emission \cite{darcy,Wimberger2003}, 
amplitude noise \cite{BP,Sadgrove2004}, 
and also to {\em large} perturbations of the kicking strength \cite{WB2004}. These counterintuitive
findings make the quantum resonances a potential source for creating directed fast atoms, 
e.g. for the study of quantum random walks \cite{qran}. However, even for narrow initial
atomic momentum distributions (i.e. those with standard deviation 
$\sigma_p <$ two-photon recoils),
true ballistic mean energy growth has only been observed  
convincingly for up to two kicks \cite{Duffy2004}.

In this letter, we present the first 
signatures of true ballistic peaks in the mean energy of a kicked atomic ensemble, with a 
relatively broad initial momentum distribution, but for an essentially classical regime.
Our seemingly counterintuitive experimental results are explained by the same technique as used
to {\em classically} describe the structure of the quantum resonance peaks of the AOKR 
\cite{Wimberger2003}. Additionally, we observe a regime in which 
an unexpected dynamical freezing effect occurs, which corresponds to 
atoms ceasing to absorb energy from the kicks.
The close proximity, in terms of pulse period, of the regimes of ballistic and frozen energy growth
promises to allow fine control of atomic velocities.

We realize the AOKR by subjecting cold Cesium atoms to a far detuned standing 
wave with spatial period $\pi/k_L$ ($k_L$ being the wave number of the kicking laser)
and pulsed with period $T$. 
Our system is described, in dimensionless units, by the Hamiltonian \cite{Graham1992}
\begin{equation}
{\cal H }(t') =\frac{p^2}{2} + k\cos(x)\sum_{t=0}^{N}
 \delta (t'-t\tau)\;,
\label{eq:ham}
\end{equation}
where $p$ is the atomic momentum in units of $2\hbar k_L$ 
(i.e. of two--photon recoils),
$x$ is the atomic position divided by $2k_L$, $t'$ is time and $t$ is an
integer which counts the kicks. 
Experimentally, we approximate $\delta$--kicks by pulses of width $\tau_p$ 
which are approximately rectangular in shape. 
We also define an effective Planck's constant $\tau=8\omega_r T$, 
where $\omega_r$ is the recoil frequency (associated with
the energy change of a Caesium atom after emission of a photon with 
$k_L=7.37\times10^6 {\rm m}^{-1}$). 
The dimensionless parameter $k$ is the kicking strength of the system. 
By exploiting the spatial periodicity of the Hamiltonian, 
we can use Bloch's theorem to reduce
the atomic dynamics along the x axis to that of a rotor on a circle, as described
in \cite{Wimberger2003}. Then,
the one-kick propagator for a given atom is 
\begin{equation} 
\Ub \;=\;e^{-{\rm i} k\cos({\hat \theta})}\;e^{-{\rm i}\tau(\Np+\beta)^2/2}, 
\label{eq:onecycle} 
\end{equation} 
where $\theta=x$mod$(2\pi)$, $\Np=-{\rm i}d/d\theta$, 
and $\beta$ is the quasimomentum (i.e., the non-integer part of $p$). To examine the limit
as $\tau \rightarrow 0$, we can use the classical version of the mapping induced by 
the Hamiltonian (\ref{eq:ham}). We define $J = \tau p$ and find
\begin{equation}
\label{eq:epsmap}
J_{t+1}=J_t+{\tilde k}\sin (\theta_{t+1})\;\;,\;\;
\theta_{t+1}=\theta_t+J_t\;,
\end{equation}
where ${\tilde k}\equiv\tau k$ is the classical stochasticity parameter of the
standard map \cite{LL1992}.
(\ref{eq:epsmap}) implicitly contains the different quasimomentum
classes which are vital to the description of quantum resonances 
\cite{Wimberger2003}. In fact, (\ref{eq:epsmap}) is equivalent to the
$\epsilon$--classical standard map developed in \cite{Wimberger2003} to
describe quantum resonance behaviour when $\tau = 2\pi l + \epsilon$ for the special 
case where $l=0$. 
Below, we will see that, in our case, the typically uniform initial 
distribution of quasimomenta \cite{Wimberger2003,darcy}
does {\em not} hinder ballistic motion at the observed resonance as 
$\tau \to 0$. \\
\begin{figure}
\centering
\includegraphics[height=4.9cm]{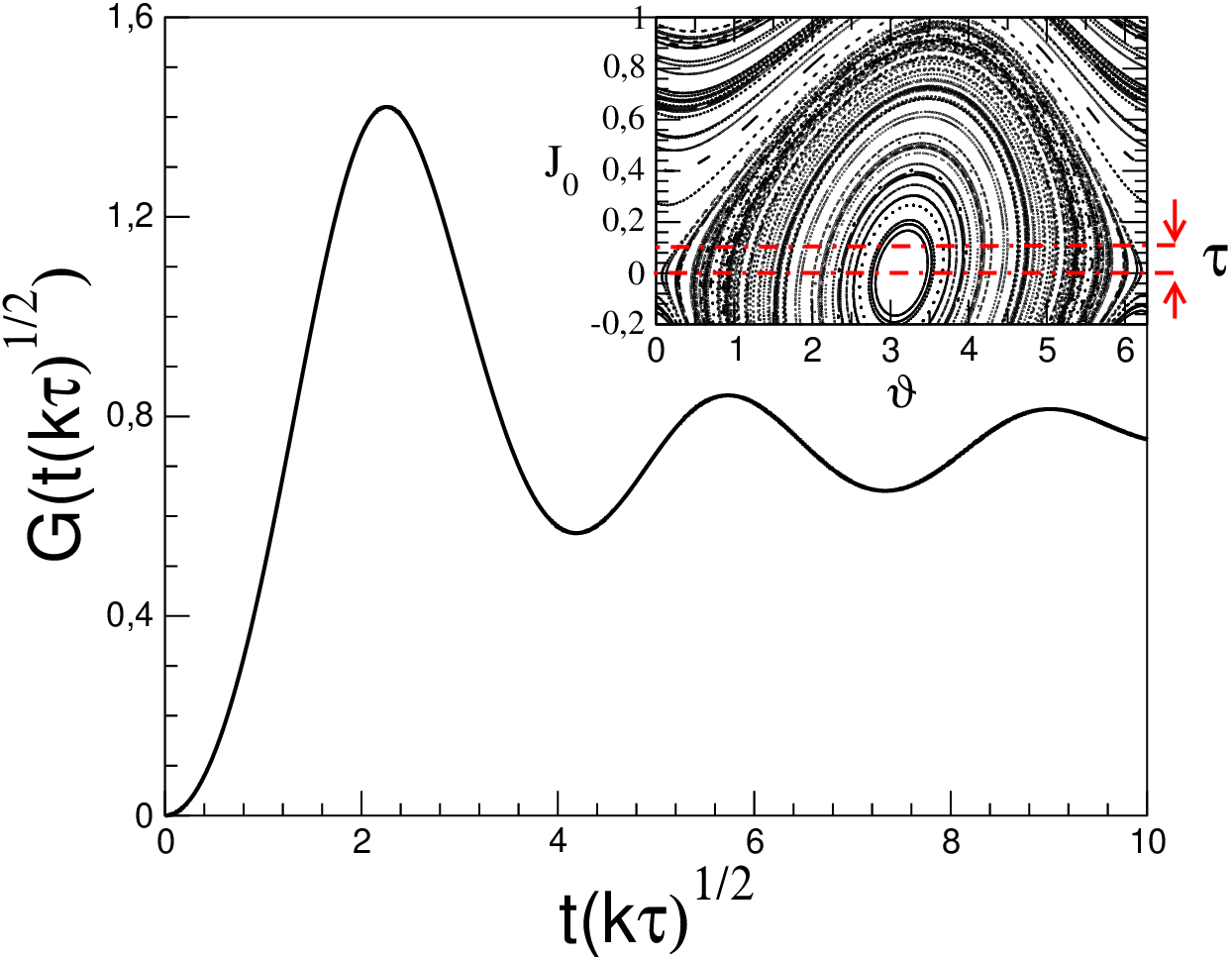}
\caption{\label{fig:phasespace} 
The function $G$ vs. $x=t(k\tau)^{1/2}$
for $J_0 \in [0,0.1)$. 
The inset shows a portion of the classical phase space of the map
(\ref{eq:epsmap}),
for $\tau = 0.1$, $k=2.5$. The dashed lines mark the position of the initial momenta, 
corresponding to a uniform distribution of quasimomenta in [0,1).
}
\end{figure}
We now compute the mean energy of an ensemble of kicked atoms from 
Eq. (\ref{eq:epsmap}).
Assuming for simplicity an initially flat distribution of 
$p\in [0,1)$ we have $J_0=\tau\beta_0$ 
with $\beta_0$ uniformly distributed in $[0,1)$. 
Since $J=\tau p$, the  mean energy of the rotor at time $t$  is given by 
$ 
\langle E_{t,\tau}\rangle= \tau^{-2}\langle(\delta J_t)^2\rangle/2\;,\;\;
\delta J_t=J_t-J_0\;. 
$
As the initial conditions in phase space populate mainly the
region close to $J_0=0$, we can
compute $\langle E_{t,\tau}\rangle$ for $\tau>0$ 
by means of the pendulum approximation \cite{LL1992}. 
The motion is described (in {\it continuous} time) 
by $H_{\rm res}=\frac12 (J')^2+\tau k\cos(\theta)$, with the
characteristic time scale for the motion in the resonant zone 
$t_{\rm res}=(k\tau)^{-1/2}$ \cite{LL1992}.
We can  formally remove $\tau$ from the Hamilton equations, 
by scaling momentum into $\overline{J}=J'/(k\tau)^{1/2}$ and 
time into $t/t_{\rm res}$. This gives
$\langle(\delta J_t)^2\rangle=\langle(J'_t-J'_0)^2\rangle\simeq 
k\tau G$,
for an  ensemble of orbits started inside the resonant zone.  
The scaling function $G(x) \equiv 
\frac{\sqrt{k}}{2\pi\sqrt{\tau}}
\int_0^{2\pi}{\rm d}\theta_0 \int_0^{\sqrt{\tau/k}}{\rm d}J_0
\overline{J}(x,\theta_0, J_0)^2 \simeq 
\frac{1}{2\pi}\int_0^{2\pi}{\rm d}\theta_0 \overline{J}(x,\theta_0, J_0=0)^2$ 
 depends on the variable $x=t(k\tau)^{1/2}$
and weakly on $k$ and $\tau$, in {\em contrast}
to the quantum resonant case studied in \cite{Wimberger2003}. 
In Fig. \ref{fig:phasespace}, $G$ is shown as a function of $x$. 
We see that $G$ tends to a level $\alpha \approx 0.7$ as $x \rightarrow \infty$.
This is because $G$ is an average over nonlinear pendulum motions with
a continuum of different periods, and therefore saturates  to a constant  value when 
the argument is larger than $\approx 1$. The dependence of $G$ on $\tau$ is negligibly
small for $\tau \lesssim 1/k$, so that practically,
$G$ can be viewed as a function of the scaling parameter $x$ alone. 
We obtain as our final result for the mean energy 
$\langle E_{t,\tau}\rangle_{\rm res} \simeq \frac{\tau k}{2\tau ^2}
G\;$.
We can now immediately derive two interesting limits.
Firstly, as we let $\tau \rightarrow 0$, the argument of $G$ becomes small,
whence $G(x) \approx x^2/2$ leading to the expression
\begin{equation}
\langle E_{t,\tau \rightarrow 0}\rangle_{\rm res} = \frac{k^2t^2}{4}\,.
\label{eq:Sandromagic1}
\end{equation}
This describes quadratic growth in mean energy which occurs as exact resonance is approached.
We note again that in the case of quantum resonances, $\epsilon$-classical theory predicts only
{\em linear} mean energy growth with kick number at resonance \cite{Wimberger2003}. 
The later linear increase is induced
by the contribution of most quasimomentum classes which lie {\em outside} the classical
resonance island for $\tau = 2\pi\ell + \epsilon$ ($\ell$ positive integer). For $\tau \rightarrow 0$,
almost all initial conditions (or quasimomenta) lie {\em within} the principal resonance island
shown in the inset of Fig. 1. 

Additionally, as we let $\tau$ grow large, $G$ saturates to the value $\alpha$ giving
\begin{equation}
\langle E_{t,\tau > 0}\rangle_{\rm res} \simeq \frac{k}{2\tau} \alpha\;,
\label{Sandromagic2}
\end{equation}
for finite $\tau>0$ and $t^2k \gg 1/\tau$. Within the stated parameter range, 
this result implies \textit{dynamical freezing} -- the ensemble's mean energy is
independent of kick number. This phenomenon is a \emph{classical} effect 
in a system with a regular phase space (see inset of Fig. \ref{fig:phasespace}).
It is distinct from dynamical localization
which is the \emph{quantum} suppression of momentum diffusion for a chaotic phase space.
Experimentally, this freezing corresponds to the cessation of energy absorption from the
kicks, similar to that which occurs at dynamical localization.
The freezing is easily explained as the averaging over all trajectories which start at momenta
close to zero, and move with different frequencies about the fixed point of the map (\ref{eq:epsmap}).

We now turn to the experimental verification of ballistic growth and 
dynamical freezing. In our experiments, we measured the mean energy of an atomic ensemble
as $\tau$ was varied and for different kick numbers.
Details of our experimental setup may be found in Ref. \cite{Sadgrove2004} and are summarized below.
Using a standard six-beam magneto-optical trap, or MOT \cite{Monroe1990}, we trap and cool
Cesium atoms typically to below $10\mu$K.
During an experiment, the trap is turned off and the atoms are subjected to pulses from
 an optical standing wave (created by a $150$mW, frequency stabilized diode laser) detuned
500 MHz from the $6S_{1/2}(F=4) \rightarrow 6P_{3/2}(F'=5)$ transition of Cesium. 
Pulse durations of $\tau_p=240$ns and $320$ns were used for our experiments with optical powers
of $P=20$mW and $30$mW respectively, and the atoms were subjected to at most $20$ kicks. For these parameters,
 the effect of spontaneous emission
is found to be negligible.
The experimental momentum distribution of the atoms is found from a 
fluoresence image of the cloud
using the standard time of flight technique \cite{Raizen,darcy,Sadgrove2004} 
and the mean kinetic
energy may be calculated from the second moment of this distribution.
In order to probe the very narrow resonance predicted above as $\tau \rightarrow 0$,
it was necessary to use very small pulse periods, leading to violations of the
$\delta$-kick assumption. 
Nonetheless, we found that
finite pulse duration effects were negligible for our results due to the relatively
low kick strengths, kick numbers and kinetic energies involved.
\begin{figure}
\centering
\includegraphics[height=5.22cm, angle=0]{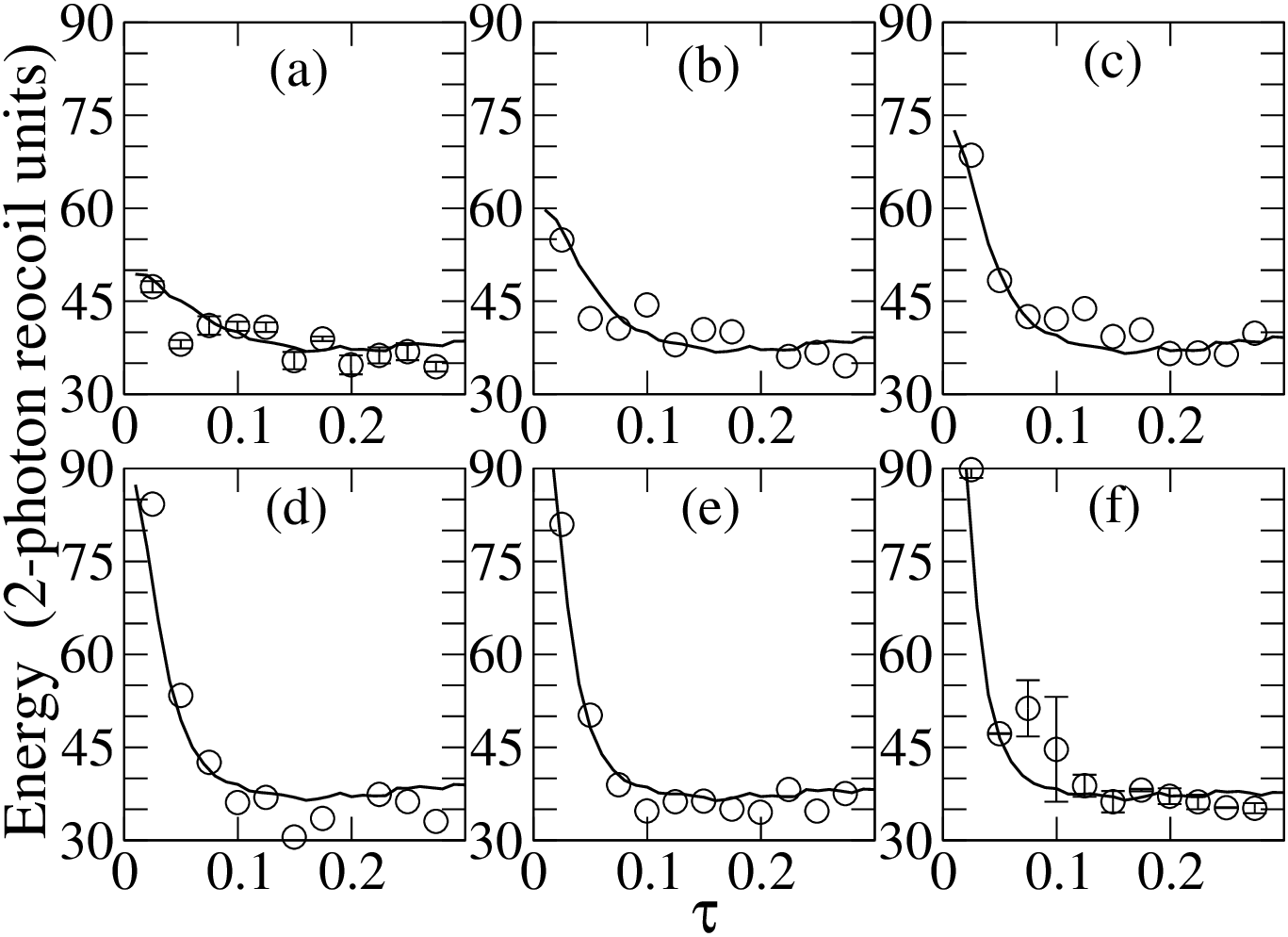}
\caption{\label{fig:k25} Results from experiments (circles) and classical simulations 
(solid line) showing energy vs. $\tau$ for $k=2.5$ and $t=3,4,5,6,7,8$ for Figs.
 (a,b,c,d,e,f) respectively. 
The error bars shown in (a) and
 (f) correspond to statistical fluctuations over three experimental shots (they 
do {\em not} take into account systematic effects such as power fluctuations of the kicking laser).}
\end{figure}
For moderate laser powers and kicking pulse durations, it is possible to estimate $k$ using
the method employed in Ref. \cite{Sadgrove2004b}. Here the difference in energies after one
and two kicks was found for values of $\tau>1$ where the energy growth is known to be quasilinear.
From the standard $\delta$-kicked rotor theory, the energy difference is $k^2/4$
\cite{LL1992}, allowing $k$ to be determined without precise knowledge of the atoms' 
initial thermal energy.
\begin{figure}
\includegraphics[height=5.22cm, angle=0]{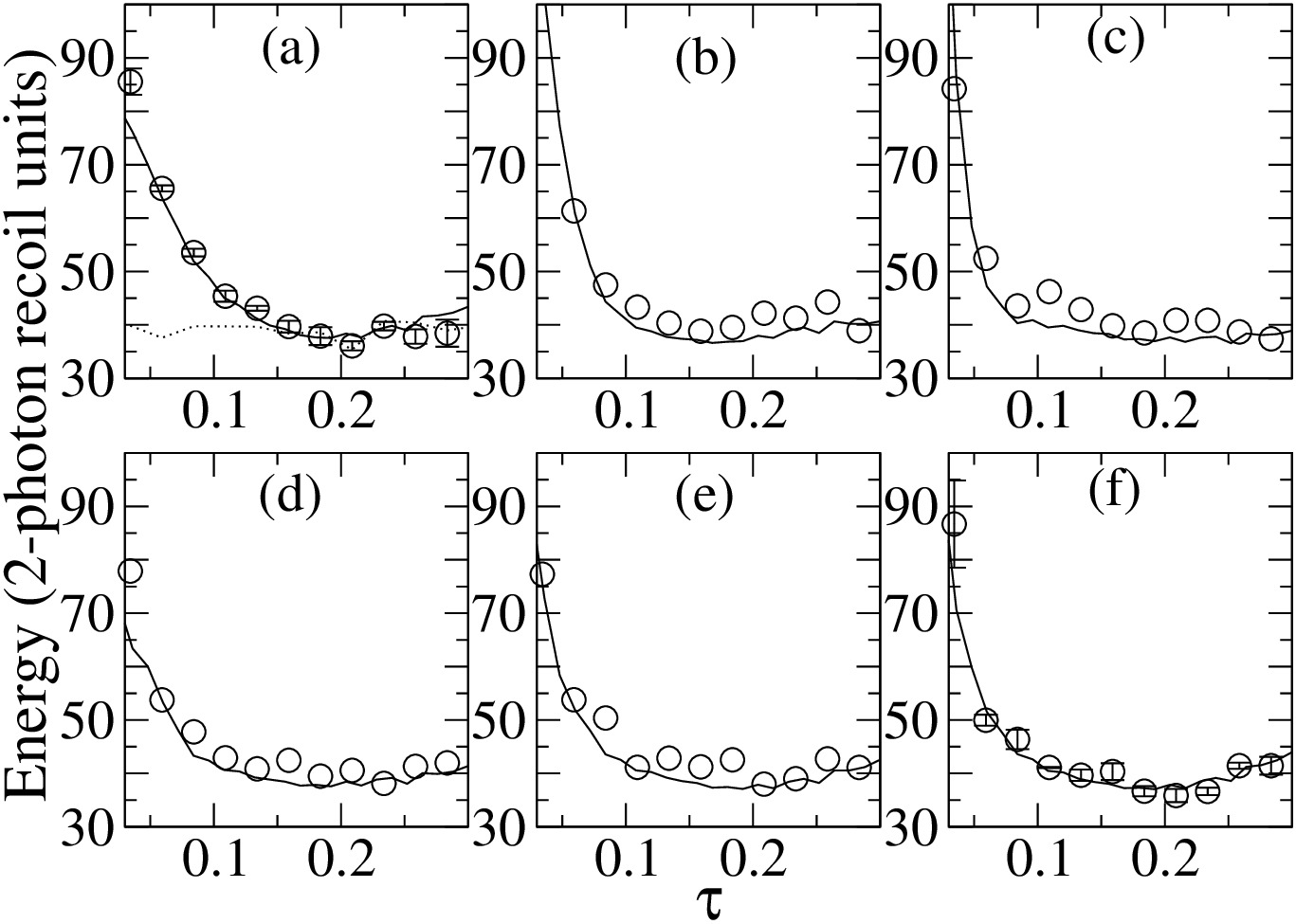}
\caption{\label{fig:k49} 
Data as in Fig.2, for $k=4.9$ and $t=3,5,7,12,16,20$ 
for Figs. (a,b,c,d,e,f)
respectively. 
The dotted line in (a) shows the measured energy after just 1 kick. 
}
\end{figure}

To predict the correct offset from zero energy of our measurements, it is still necessary to 
calculate the standard deviation $\sigma_p$ of the initial experimental momentum distribution.
The calculation of $\sigma_p$ requires truncation
of the wings of the experimental momentum distribution at some momentum
so that the second moment calculation is not ruined by noise in the
wings. Since the initial momentum distribution has considerable non--Gaussian wings,
 we inevitably underestimate the value of $\sigma_p$ using this method.
We estimate this systematic error to be less 
than $20\%$ or $1.5$ two-photon recoils for the experimental results presented here. 
Fig. \ref{fig:k25} shows measured energies (circles) and classical data for a measured
value of $k=2.5\pm 0.1$, and various kick numbers. The classical results are
obtained from the dynamics generated by (\ref{eq:epsmap}) for a Gaussian initial ensemble
of 25000 momenta $p=J/\tau$, with 200 uniformly distributed angles each.
We have taken $\sigma_p=8.4$ for our simulations, and the results can be compared with the predictions of
Eqs. (\ref{eq:Sandromagic1}) and (\ref{Sandromagic2}) if the initial ensemble's energy is added to the analytic results
(derived above for $p \in [0,1)$). Allowing for experimental uncertainties, excellent agreement is found
between the measurements and our classical theory. In 
particular, a ballistic resonance peak is seen to exist and, although 
the vanishing resonant kicking period 
itself cannot be probed, the trend towards quadratic growth is clear in the data. The resonance 
is very narrow -- much more so than the quantum resonances for the same parameters. Specifically,
the half width of the $\tau \rightarrow 0$ peak after five kicks is $\sim 0.05$ whereas that for a
quantum resonance peak for the same value of $k$ would be $\sim 0.4$ \cite{Wimberger2003}.
The extremely fast compression of the
resonance peak is characteristic of a quantum resonance which would be observed for a purely
plane wave initial condition with momentum $p=0$ at $\tau = 4\pi$, for instance. Hence, the 
peak width observed here shrinks even faster than at the quantum resonances observed in \cite{darcy}, for which
a sub--Fourier peak width $\propto 1/t^2$ was predicted \cite{Wimberger2003}. This makes the resonance
at vanishing kicking period a sensitive phenomenon with the potential to be useful for making
high-precision measurements.
\begin{figure}
\includegraphics[height=5.0cm, angle=0]{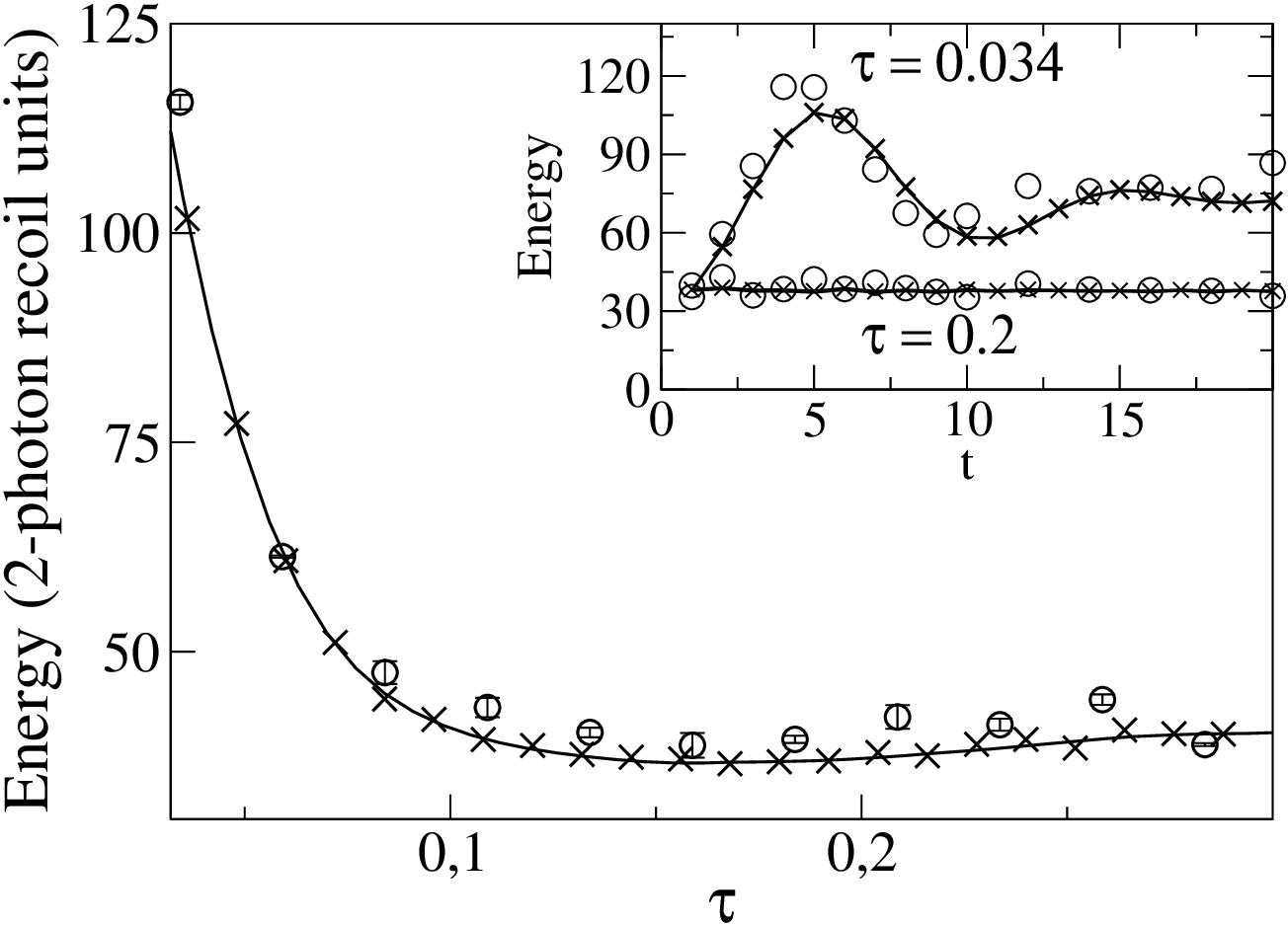}
\caption{\label{fig:qmcompare} Mean energy vs. $\tau$ 
from experiments (circles), classical
(crosses) and quantum simulations (using a Gaussian initial ensemble
centered at $p=0$; solid line) for $k=4.9$, $\sigma_p=8$ and $t=5$.
Note the excellent agreement between experimental data and the 
simulations. The inset shows
how the energy varies with $t$ for $\tau=0.2$ and the 
smallest measurable period $\tau=0.034$ (corresponding to 
$T=0.33\mu$s). We also note the immediate onset of
the freezing effect for $\tau = 0.2$, whilst dynamical localization 
would set in on a time scale much larger than 1 kick for the present kick 
strength \cite{Shepelyansky1987}.
}
\end{figure}
The results in Fig.~\ref{fig:k25} also demonstrate that dynamical freezing, i.e.,
zero net energy gain, is occurring for $\tau > 0.1$ after multiple kicks. To 
observe this localization effect more convincingly, we chose a larger value of $k$ and 
measured the mean energy of our atoms for up to 20 kicks. The results, shown in Fig. 
\ref{fig:k49}, are for a measured value of $k=4.9 \pm 0.2 $, with $\sigma_p$ taken to 
be $8$ for the simulations. After 20 kicks, for $\tau > 0.1$, the mean energy 
of the atoms has not risen above the one kick level. Again, excellent agreement is seen 
between experimental results and classical data.
As a final test we have compared our data with quantum simulations for 
$k=4.9$, $\sigma_p=8$. As shown in Fig. \ref{fig:qmcompare}, 
the numerics and experimental measurements 
agree almost perfectly within the estimated errors. On closer inspection, 
a hint that the classical
approximation fails for large $\tau$ can be found in Fig. 4, where
the classical data starts to deviate (very slightly) from the quantum data for larger
$\tau \geq 1/k$. Above that threshold, the standard map enters the critical regime,
which mainly affects the small subclass of extremely large momenta in our
initial ensemble, lying outside
the principal resonance island. Stronger deviations are expected in this 
region for very large interaction times ($t\gg 10$) \cite{Wimberger2003}.
The inset in Fig.~\ref{fig:qmcompare} demonstrates the regimes of 
near resonant and off-resonant (frozen) energy growth
more explicitly by plotting the mean energy against kick number. The curve for
$\tau=0.034$ closely resembles the structure of the ``quasi'' scaling function
$G$ from Fig. 1, showing both the initial ballistic growth and saturation
after $t\geq 2/\sqrt{k\tau}\simeq 5$ kicks. 

In conclusion, we have experimentally observed and theoretically explained the
occurrence of ballistic, resonance like transport in the limit of vanishing kicking period
for a broad initial ensemble of atomic momenta. 
The {\em quadratic} energy growth at this resonance has
been verified in comparison with the {\em linear} growth 
of mean energy found at quantum resonance
for comparable initial conditions \cite{darcy,Sadgrove2004}.
Away from the ballistic peak, a dynamical freezing effect, 
as a consequence of the underlying regular
classical dynamics, has been observed, corresponding to zero energy-growth 
rates for up to 20 kicks in our experiment.
Both of these phenomena, realized here for the first time, are of great
interest for experimentally controlling fast (ballistic) and slow 
(frozen) atomic velocities.


\end{document}